\documentclass[journal=nalefd,manuscript=article,hyperref]{achemso}

\usepackage[version=3]{mhchem}

\title{Directional and dynamic modulation of the optical emission of an individual \ce{GaAs} nanowire using surface acoustic waves}

\author{J\"org B. Kinzel}
\affiliation[Univ. Augsburg]{Lehrstuhl f\"{u}r Experimentalphysik 1 and Augsburg Centre for Innovative Technologies (ACIT), Universit\"{a}t Augsburg, Universit\"{a}tsstr. 1, 86159 Augsburg, Germany}
\author{Daniel Rudolph}
\affiliation[TU M\"unchen]{Walter Schottky Institut and Physik Department, Technische Universit\"{a}t M\"{u}nchen, Am Coulombwall, 85748 Garching, Germany}
\author{Max Bichler}
\affiliation[TU M\"unchen]{Walter Schottky Institut and Physik Department, Technische Universit\"{a}t M\"{u}nchen, Am Coulombwall, 85748 Garching, Germany}
\author{Gerhard Abstreiter}
\affiliation[TU M\"unchen]{Walter Schottky Institut and Physik Department, Technische Universit\"{a}t M\"{u}nchen, Am Coulombwall, 85748 Garching, Germany}
\author{Jonathan J. Finley}
\affiliation[TU M\"unchen]{Walter Schottky Institut and Physik Department, Technische Universit\"{a}t M\"{u}nchen, Am Coulombwall, 85748 Garching, Germany}
\author{Gregor Koblm\"{u}ller}
\affiliation[TU M\"unchen]{Walter Schottky Institut and Physik Department, Technische Universit\"{a}t M\"{u}nchen, Am Coulombwall, 85748 Garching, Germany}
\author{Achim Wixforth}
\affiliation[Univ. Augsburg]{Lehrstuhl f\"{u}r Experimentalphysik 1 and Augsburg Centre for Innovative Technologies (ACIT), Universit\"{a}t Augsburg, Universit\"{a}tsstr. 1, 86159 Augsburg, Germany}
\author{Hubert J. Krenner}\email{hubert.krenner@physik.uni-augsburg.de}
\affiliation[Univ. Augsburg]{Lehrstuhl f\"{u}r Experimentalphysik 1 and Augsburg Centre for Innovative Technologies (ACIT), Universit\"{a}t Augsburg, Universit\"{a}tsstr. 1, 86159 Augsburg, Germany}

\begin{document}
\begin{abstract}
We report on optical experiments performed on individual $\rm GaAs$ nanowires and the manipulation of their temporal emission characteristics using a surface acoustic wave. We find a pronounced, characteristic suppression of the emission intensity for the surface acoustic wave propagation aligned with the axis of the nanowire. Furthermore, we demonstrate that this quenching is dynamical as it shows a pronounced modulation as the local phase of the surface acoustic wave is tuned. These effects are strongly reduced for a SAW applied in the direction perpendicular to the axis of the nanowire due to their inherent one-dimensional geometry. We resolve a fully dynamic modulation of the nanowire emission up to $678$ MHz not limited by the physical properties of the nanowires.
\end{abstract}

Semiconductor nanowires (NWs) offer a particularly promising platform to combine both electrostatic \cite{Bjork:05a,*Fasth:05a,*Pfund:06} and optically active heterostructure quantum dots (QDs) \cite{Bjork:02,*Weert:09a,*Kouwen:10a} in an inherently scalable architecture for the implementation of electro-optical \cite{Gywat:09:book,*Hanson:07} spin-based quantum information schemes. Furthermore, these types of nanostructures allow for confinement of both electronic and photonic degrees of freedom\cite{Claudon:10a} providing a direct interconnect between localized spin and flying photonic qubits. To fully exploit the advantages of isolated electron spins in electrostatic QDs and their conversion into photons in a heterostructure QD on \emph{the same NW} a non-invasive, fast bus is required. This bus could be implemented using spin-preserving acoustoelectric charge conveyance\cite{Sogawa:01a,*Stotz:05}  mediated by high-frequency surface acoustic waves (SAWs). SAWs have a long-standing history as a versatile tool to dynamically control and manipulate embedded optically active nanosystems such as quantum wells \cite{Rocke:97}, quantum wires \cite{Alsina:02a} and quantum dots \cite{Gell:08,*Metcalfe:10} and columnar quantum posts \cite{Voelk:10b} with frequencies ranging from tens of MHz up to several GHz. In these experiments acoustic charge conveyance of charges, spins and dissociated excitons, inherently sequential electron-hole injection and modulation of the emission wavelength of these planar semiconductor heterostructures have been observed. In contrast, reports on SAW control of isolated one-dimensional nanosystems such as nanotubes and NWs are so far limited to electric transport experiments \cite{Ebbecke:08,*Roddaro:10} in which only one carrier species is probed.\\

Here we demonstrate \emph{dynamic} control of the emission of single \ce{GaAs} NWs employing high frequency SAWs. We find a pronounced directionality of this process as its efficiency is strongly enhanced for SAWs propagating along the NW axis. This observation is in excellent agreement with a dissociation process of photogenarated electron-hole pairs by the piezoelectric fields generated by the SAW. This process is inhibited for SAWs propagating perpendicular to the NW axis since their diameter is more than one order of magnitude smaller than the wavelength of the SAW. This is in strong contrast to embedded low-dimensional semiconductor nanostructures which are typically surrounded and coupled to systems of higher dimensionality (quantum wells and bulk).
\\

\begin{figure}[htb]
    \begin{center}
 \includegraphics[width=0.5\columnwidth]{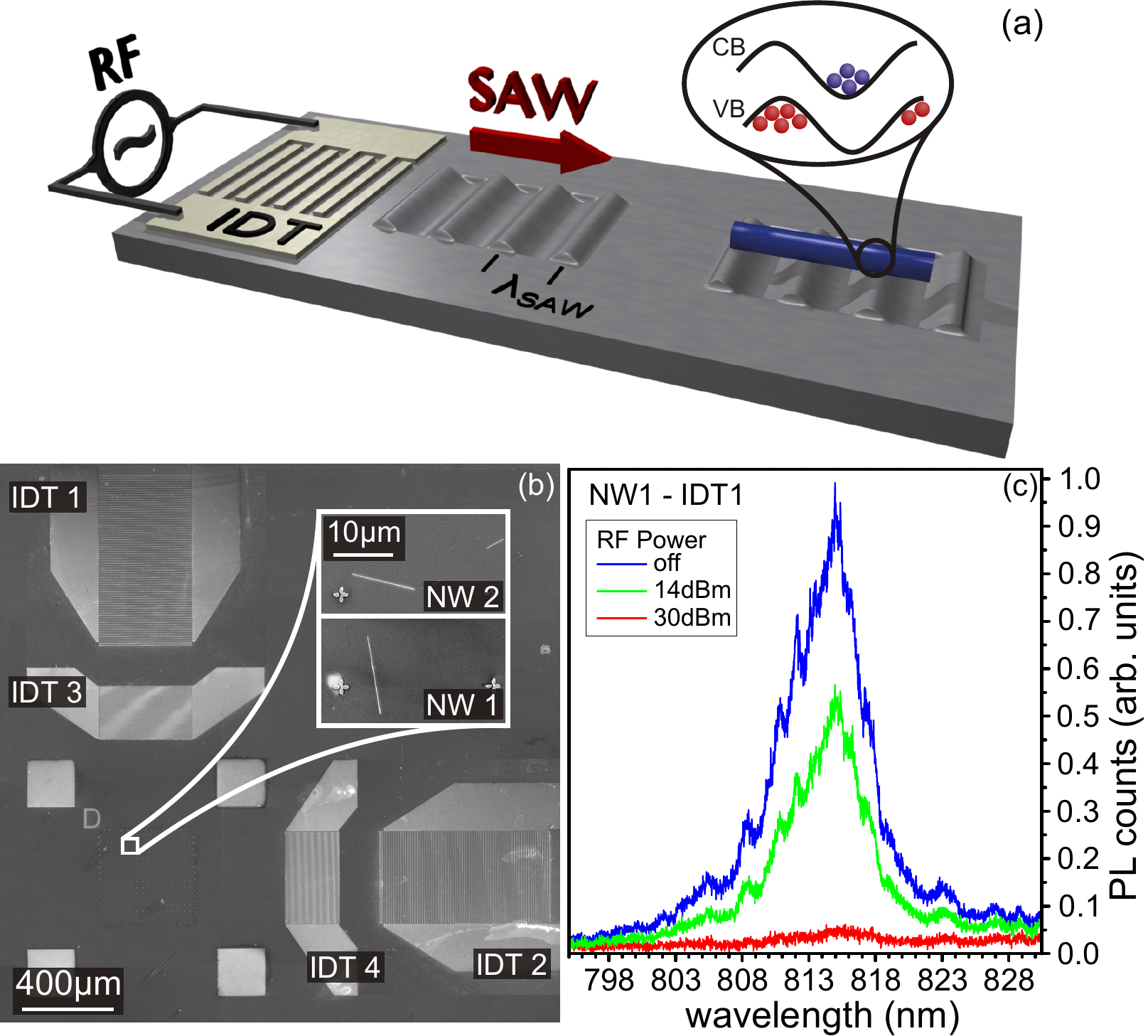}
        \caption{(a) Schematic of the device consisting of an IDT to generate a SAW. The SAW induces a Type-II bandedge modulation in the NW which leads to a dissociation of electrons and holes. (b) SEM images of the SAW-device and the two NWs studied. (c) Emission spectra of {\bf NW1} without (blue) and a 210 MHz SAW generated by $P_{\rm RF} = 14$ dBm (green) and 30 dBm (red) applied to {\bf IDT3} showing pronounced suppression with increased RF power. }
        \label{fig:1}
    \end{center}
\end{figure}

The NWs studied were fabricated by molecular beam epitaxy (MBE) using an autocatalytic growth process on Si (111) substrates covered by
an ultra-thin amorphous $\rm SiO_ x$ nucleation mask \cite{Morral:08a,*Koblmueller:10a}. These NWs have typical lengths
of $l_{\rm NW}> 15~\mathrm{\mu m}$ and diameters of $d_{\rm NW}\sim50-120$ nm with predominantly zinc blende crystal structure
as confirmed by X-ray diffraction, transmission electron microscopy and Raman spectroscopy \cite{Spirkoska:09}. We prepared
a hybrid SAW device to study individual NWs, a schematic and a scanning electron micrograph (SEM) of which is shown in \ref{fig:1} (a) and (b), respectively. First, we realize a SAW chip by defining interdigital transducers (IDTs) labeled {\bf IDT1}--{\bf IDT4} on a $128^{\circ}$ XY-cut $\rm LiNbO_3$ substrate using electron beam
lithography and a lift-of process. The periodicities chosen for {\bf IDT1}/{\bf IDT2} and {\bf IDT3}/{\bf IDT4} allow for the excitation of SAWs of wavelengths $\lambda_{1,2} = 17.5~\rm \mu m$ and $\lambda_{3,4} = 5.5 ~\rm \mu m$, respectively. At low temperatures we determined the corresponding resonance frequencies to be $f_1=$ 210 MHz, $f_2=$ 222 MHz and $f_3=f_4=678$ MHz, respectively. The two pairs are rotated by $90^{\circ}$, and, thus on this type of substrate SAWs with a propagation direction $\hat{k}_{\rm SAW}$~ can be excited along the two perpendicular crystal directions X ({\bf IDT1}, {\bf IDT3}) and $128^{\rm o}$ off Y ({\bf IDT2}, {\bf IDT4}). These two propagation directions are in the following referred to as the down and left directions, respectively.
In a second step we dispersed a low concentration of NWs which are randomly distributed and oriented on the SAW chip. In the intersection region of the propagation paths of the two IDT pairs, we identify individual NWs using SEM. Here we report on experiments performed on two individual NWs labeled {\bf NW1} and {\bf NW2} in \ref{fig:1} (b). These two NWs have been selected due to the relative orientation of their axes ($\hat{a}_{\rm NW}$) with respect to the propagation directions of SAWs ($\hat{k}_{\rm SAW}$). SAWs generated along the up (left) direction are aligned (perpendicular) to the axis of {\bf NW1} while the situation is completely reversed for {\bf NW2} as shown in \ref{fig:1} (b). A detailed description on the MBE growth, the used growth parameters and details on the sample fabrication are summarized in the Supporting Information.\\

We study their emission using low temperature ($T\sim$5 K) micro-photoluminescence ($\rm \mu$-PL). Electron-hole pairs were photogenerated using either a diode laser emitting $\tau_{laser}< 100 \rm ~ps$ long pulses ($\lambda = 661$ nm) or a continuous wave HeNe laser ($\lambda = 632.8$ nm) which we focused to a $\sim 3 \mathrm{\mu m}$ diameter spot using a $50\times$ microscope objective. The NW emission was collected via the same objective and dispersed using a 0.5 m grating monochromator with an overall resolution $<0.15\mathrm{~meV}$. The signal was detected either time-integrated by a liquid \ce{N2} cooled multi-channel Si-CCD camera or a single channel Si-single photon counting module. The latter provides a temporal resolution of $<300~\mathrm{ps}$ and is used for time-correlated single photon counting (TCSPC).
In all experiments we used 900 ns long SAW pulses with a repetition rate of 100 kHz which we either synchronized to the train of laser pulses ($f_{laser}=80~{\rm MHz}$) or used as a timing reference for TCSPC. Moreover, we can lock time excitation laser pulses to a stable, arbitrary tunable phase of the SAW by setting $f_{laser}$ such that $(n\cdot f_{laser}= f_{\rm SAW}$, where  $n$ integer)\cite{Voelk:11a}. Both techniques enable us to obtain high temporal resolution to resolve dynamical processes over timescales defined by the SAW and resolve their \emph{full} phase information. The Supporting Information accompanying this paper contains a detailed description of implementation these synchronization schemes.\\
Typical emission spectra of {\bf NW1} recorded under weak optical pump powers $<3~\mathrm{\mu W}$ are presented in \ref{fig:1} (c) without and with an 210 MHz SAW generated by {\bf IDT1} with $P_{\rm RF} =+14$ dBm and +30 dBm parallel to the $\hat{a}_{\rm NW}$. For no SAW applied we observe a strong PL signal from {\bf NW1} at $\lambda= 818\pm 4$ nm. \footnote{The substructure of the emission peak arises form a weak modulation of the sensitivity of our detector in this spectral range due to etaloning.} As the $P_{\rm RF}$ is increased to $P_{\rm RF} =+14$ dBm we observe a pronounced suppression of the NW emission which is almost completely quenched at  $P_{\rm RF} =+30$ dBm. This characteristic quenching of the PL signal arises from a dissociation of photogenerated excitons by a Type-II band edge modulation along the SAW propagation direction as shown in \ref{fig:1} (a). This effect has been previously observed for embedded semiconductor heterostructures \cite{Rocke:97,Alsina:02a,Voelk:10b}. In contrast to these structures, the NWs studied in the present work resemble isolated one-dimensional systems. Thus, we expect any SAW-driven process to become highly directional depending on the relative orientation of the NW axis and the SAW propagation direction.

\begin{figure}[htb]
    \begin{center}
\includegraphics[width=0.5\columnwidth]{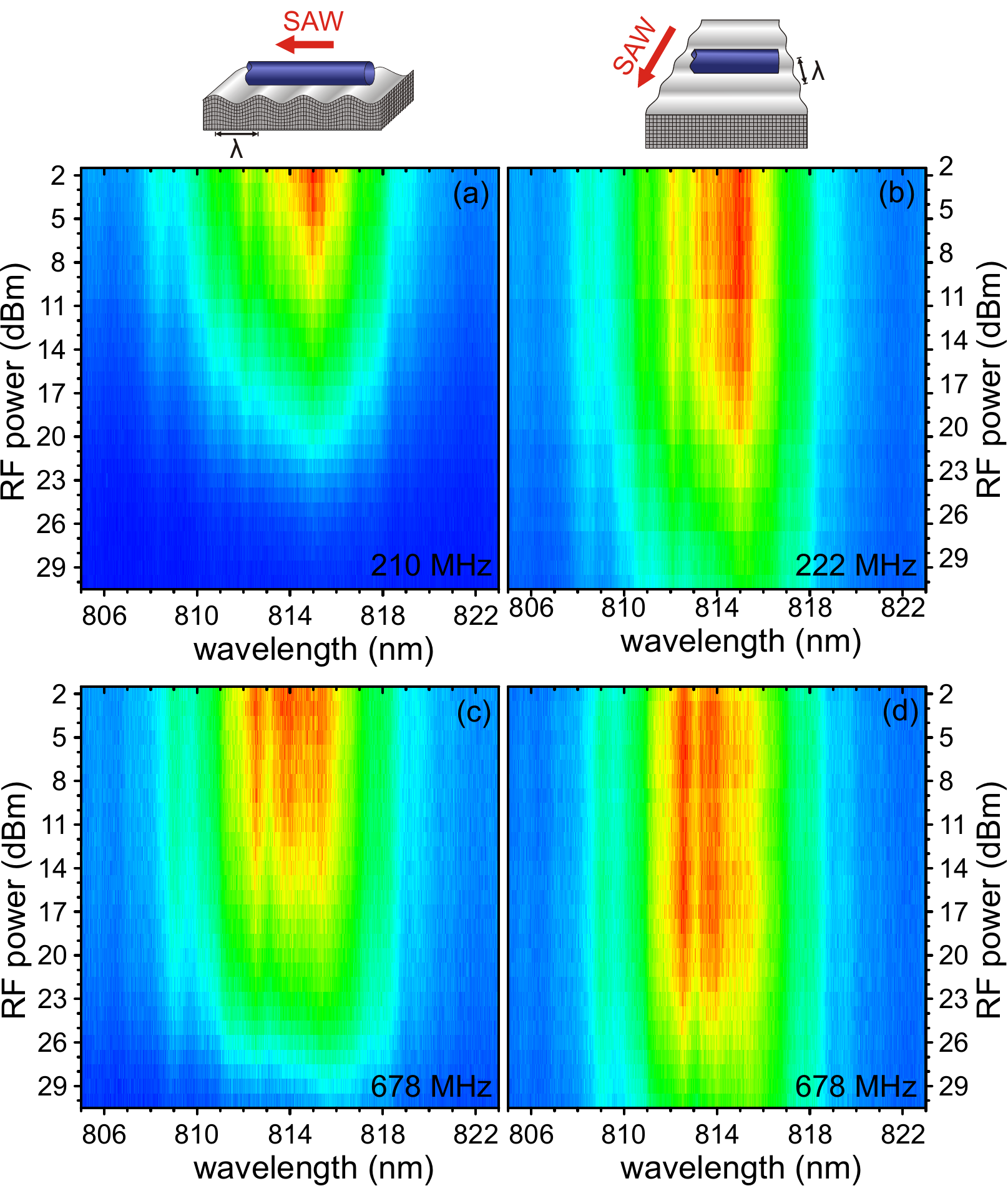}
        \caption{Emission of {\bf NW1} for different SAW propagation directions [schematics] and frequencies  [210/222 MHz (a) and (b); 678 MHz (c) and (d)] as a function of RF power. For both frequencies the PL signal quenching with increasing $P_{\rm RF}$ is enhanced for the SAW propagating along the NW axis.}
        \label{fig:2}
    \end{center}
\end{figure}
We investigated the SAW driven quenching of the NW emission in detail as a function of the SAW power, frequency and relative orientation of $\hat{k}_{\rm SAW}$~with respect to $\hat{a}_{\rm NW}$. The results of these experiments performed on {\bf NW1} are presented in \ref{fig:2}. We plot the detected PL spectra normalized to the unperturbed emission in false color representation as a function of the applied RF power for low (high) frequency SAWs in the upper and lower upper panels, respectively. The relative orientation is $\hat{k}_{\rm SAW}$$\parallel$$\hat{a}_{\rm NW}$~and $\hat{k}_{\rm SAW}$$\perp$$\hat{a}_{\rm NW}$~in the left and right panels as indicated by the schematics. In experiment we observe similar behavior of the NW emission for both SAW frequencies as we increase $P_{\rm RF}$. For $\hat{k}_{\rm SAW}$$\parallel$$\hat{a}_{\rm NW}$~we observe a rapid and very pronounced quenching of the PL signal setting in at $P_{\rm RF}= +3\mathrm {~dBm}$ and $P_{\rm RF}= +17\mathrm {~dBm}$ for the low and high SAW frequencies. This dependence is in excellent agreement with a more efficient exciton dissociation within the Type-II band edge modulation with increasing SAW amplitude i.e. RF power. The reduced efficiency for the higher frequency is mainly attributed to a reduced coupling efficiency of the applied RF power to SAWs compared to the lower frequency. In strong contrast to planar, embedded nanostructures, we find that for both SAW frequencies the onset of the PL suppression is shifted significantly to $P_{\rm RF}= +8\mathrm {~dBm}$ and $P_{\rm RF}= +23\mathrm {~dBm}$ for $\hat{k}_{\rm SAW}$$\perp$$\hat{a}_{\rm NW}$~compared to the parallel configuration. This experimentally observed, pronounced directionality of the emission suppression can be readily understood by taking into account the geometry of the inherently one-dimensional NWs. Clearly, the diameter of the NW is significantly smaller than the wavelength of the SAW $ d_{\rm NW} \ll \lambda_{\rm SAW}$ whilst $ l_{\rm NW} > \lambda_{\rm SAW}$. Since $\lambda_{\rm SAW}$ defines the lengthscale on which carrier separation and exciton dissociation occur [c.f. \ref{fig:1} (a)], we expect the efficiency of this SAW-driven process to be significantly reduced (enhanced) for a SAW propagating perpendicular (parallel) to the $\hat{a}_{\rm NW}$~which is nicely observed in our experimental data shown in \ref{fig:2}. In addition we note that we do not resolve spectral shifts in our data, as observed e.g. for the narrow PL lines of QDs \cite{Gell:08,*Metcalfe:10,Voelk:10b} at large SAW amplitudes, due to the relatively broad emission peak of our NWs.  \\

\begin{figure}[htb]
    \begin{center}
     \includegraphics[width=0.5\columnwidth]{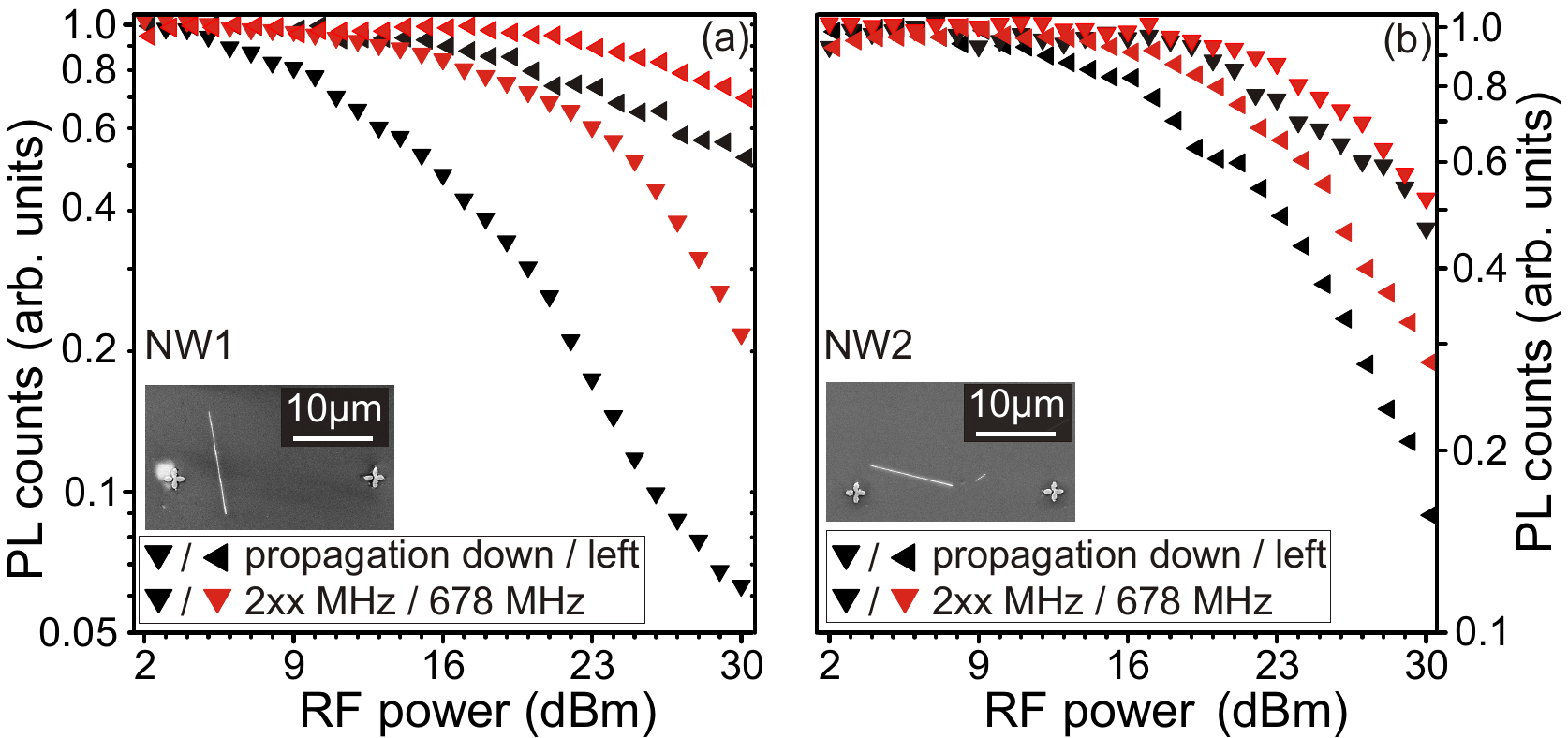}
        \caption{Comparison of the SAW-induced suppression of the integrated PL intensity for {\bf NW1} (a) and {\bf NW2} (b) for SAWs propagating down (down triangles) or left (left triangles). For both NWs we observe the most pronounced suppression for $\hat{k}_{\rm SAW}$$\parallel$$\hat{a}_{\rm NW}$~i.e. down triangles {\bf NW1} and left triangles {\bf NW2} independent on the SAW frequency.}
        \label{fig:3}
    \end{center}
\end{figure}

However, we have to exclude that the observed directionality does not arise from the anisotropic piezoelectric properties for SAW excitation in the down and left directions for the particular substrates used for our SAW device. As a first characterization we confirmed the comparable, efficient SAW generation both in the down and left directions by studying the diffraction efficiency of the SAW-induced amplitude grating \cite{Ruppert:10}. To fully confirm that the observed directionality arises from a more efficient exciton dissociation for $\hat{k}_{\rm SAW}$$\parallel$$\hat{a}_{\rm NW}$~compared to $\hat{k}_{\rm SAW}$$\perp$$\hat{a}_{\rm NW}$~we performed identical experiments as presented for {\bf NW1} in \ref{fig:2} on {\bf NW2} which is rotated by $\sim$90$\rm^o$ with respect to {\bf NW1}. Thus, we expect the quenching of the emission of {\bf NW2} to be more efficient for a left propagating SAW compared to a down propagating SAW. In \ref{fig:3} we compare the integrated PL intensities detected from {\bf NW1} (a) and {\bf NW2} (b) as a function of $P_{\rm RF}$. The presented experimental data clearly confirms that the emission of {\bf NW2} is more efficiently quenched by a SAW propagating to the left i.e. $\hat{k}_{\rm SAW}$$\parallel$$\hat{a}_{\rm NW}$~(left triangles) than by a SAW propagating down $\hat{k}_{\rm SAW}$$\perp$$\hat{a}_{\rm NW}$~(down triangles) for both SAW frequencies studied. This directionality is completely reversed compared to {\bf NW1} proving that the observed anisotropy indeed arises only from the relative orientation of $\hat{a}_{\rm NW}$~and $\hat{k}_{\rm SAW}$.\\

\begin{figure}[htb]
    \begin{center}
   \includegraphics[width=0.5\columnwidth]{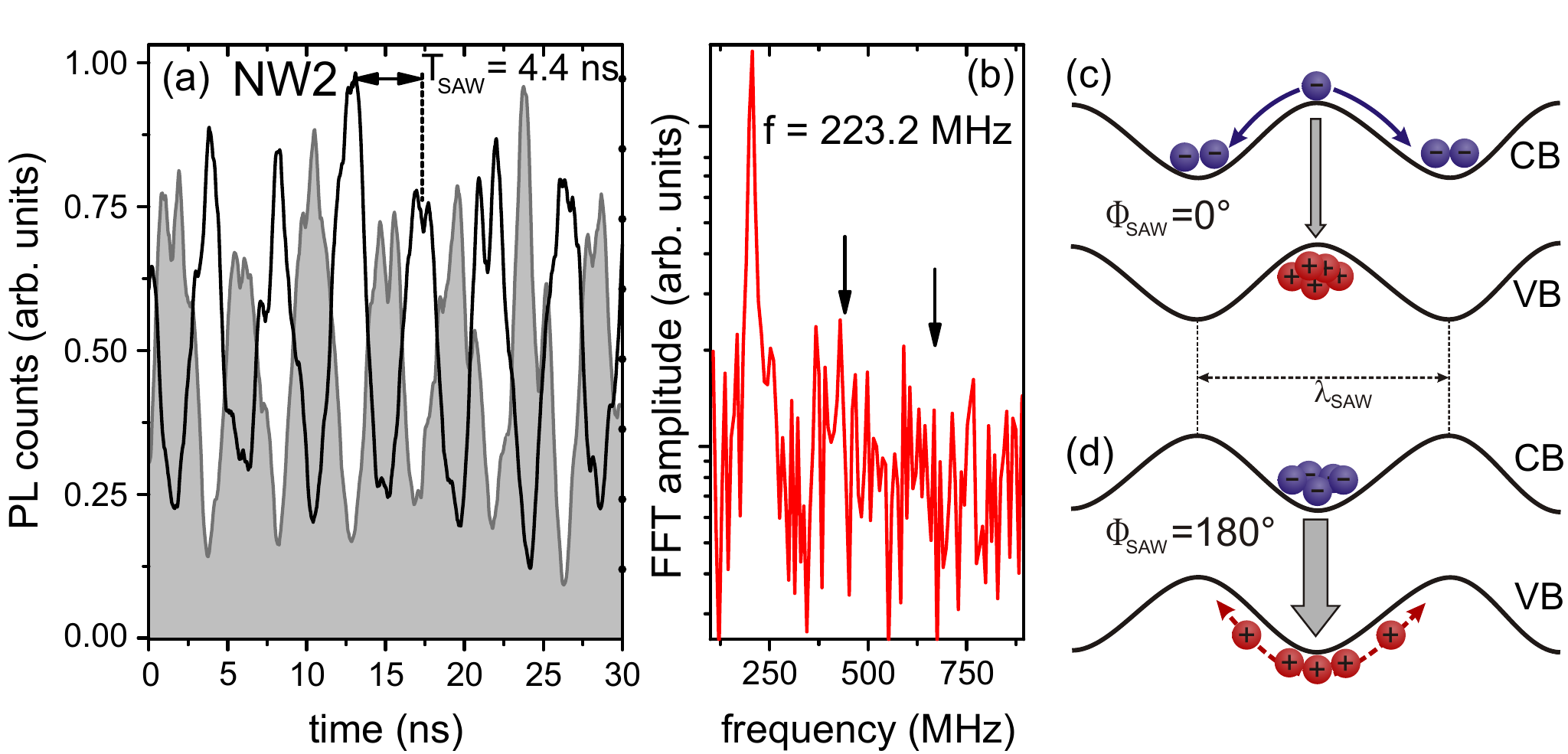}
    \caption{(a) TCSPC histogram recorded of the emission of {\bf NW2} modulated by 222 MHz SAWs with a relative phase shift of $180^{\rm o}$. (b) Fourier transform of (a) showing the modulation with the frequency of the SAW.}
        \label{fig:5}
    \end{center}
\end{figure}

Up to now we neglected that this process itself is dynamic on a timescale defined by the period of the SAW $(T_{\rm SAW})$. In order to resolve this fast dynamic modulation of the NW emission we perform TCSPC spectroscopy on the emission of {\bf NW2}. In this experiment we measure the temporal correlation between the detected NW emission excited by a cw laser and the SAW. For low RF power $P_{\rm RF} <+10{\rm~dBm}$ applied at $f_{\rm SAW}= 222 {~\rm MHz}$ to {\bf IDT2} no modification of the total emission signal and, thus, no temporal correlations are observed. However, when we increase the RF power to $P_{\rm RF} =+27 {\rm~dBm}$, we find a clear modulation of the TCSPC signal as shown in the solid line in \ref{fig:5} (a). The period of the observed modualtion $4.4 \pm0.3{\rm ~ns}$ agrees well with the SAW period $T_{\rm SAW}=1/f_{\rm SAW}=4.4 {\rm ~ns}$ providing direct evidence for a \emph{dynamic} acoustoelectric modulation of the NW emission on a nanosecond timescale. Moreover, only modulation with the fundamental period of the SAW $f = 223.2 {\rm ~MHz}$ and no contribution from higher harmonics (marked by arrows) are observed in the Fourier transform of the TCSPC signal [c.f. \ref{fig:5} (b)]. This modulation is characteristic for a \emph{diffusive} exciton dissociation process in the SAW-induced band edge modulation schematically shown in \ref{fig:5} (c) and (d): At $\phi_{\rm SAW}=0^{\rm o}$ [\ref{fig:5} (c)] holes are generated in an effective potential that is a stable maximum of the VB whilst electrons are located at a unstable maximum in the CB. This situation is reversed at $\phi_{\rm SAW}=180^{\rm o }$ at which electrons are generated at a stable minimum in the CB and holes at a unstable minimum in the VB [\ref{fig:5} (d)]. Due to their higher mobility electrons diffuse rapidly from the point of generation for $\phi_{\rm SAW}=0^{\rm o }$. This effect leads to a strong reduction of the electron density  $\phi_{\rm SAW}=0^{\rm o }$ as indicated in \ref{fig:5} and consequently gives rise to the observed pronounced reduction of the PL emission. At $\phi_{\rm SAW}=180^{\rm o }$ electrons are captured at the stable minimum in the CB while at the same time hole diffusion is limited due to the higher effective mass and smaller mobility of this carrier species. Thus, the PL emission is only weakly suppressed at this local phase of the SAW corresponding to the maxima in the TCSPC signal. We note that similar observations have been reported for planar, fully embedded semiconductor heterostructures \cite{Alsina:02a,Alsina:03} confirming our interpretation. This SAW-driven modulation requires that the PL decay time is significantly faster than $T_{\rm SAW}/2$\cite{Voelk:11a}. This condition would not be met in mixed crystal phase (zinc blende/wurtzite)NWs for which PL decay times can exceed several nanosconds\cite{Spirkoska:09}. To further support this model we performed a control experiment in which we shifted the phase of the SAW by $180^{\rm o}$ with respect to the rising edge of the pulse used a the reference in the TCSPC experiment and plot the detected modulated signal in \ref{fig:5} (a). We find this histogram (gray shaded) offset by $T_{\rm SAW}/2$ with respect to the original trace (line) corresponding to the externally set shift of the local SAW phase.\\
\begin{figure}[htb]
    \begin{center}
    \includegraphics[width=0.5\columnwidth]{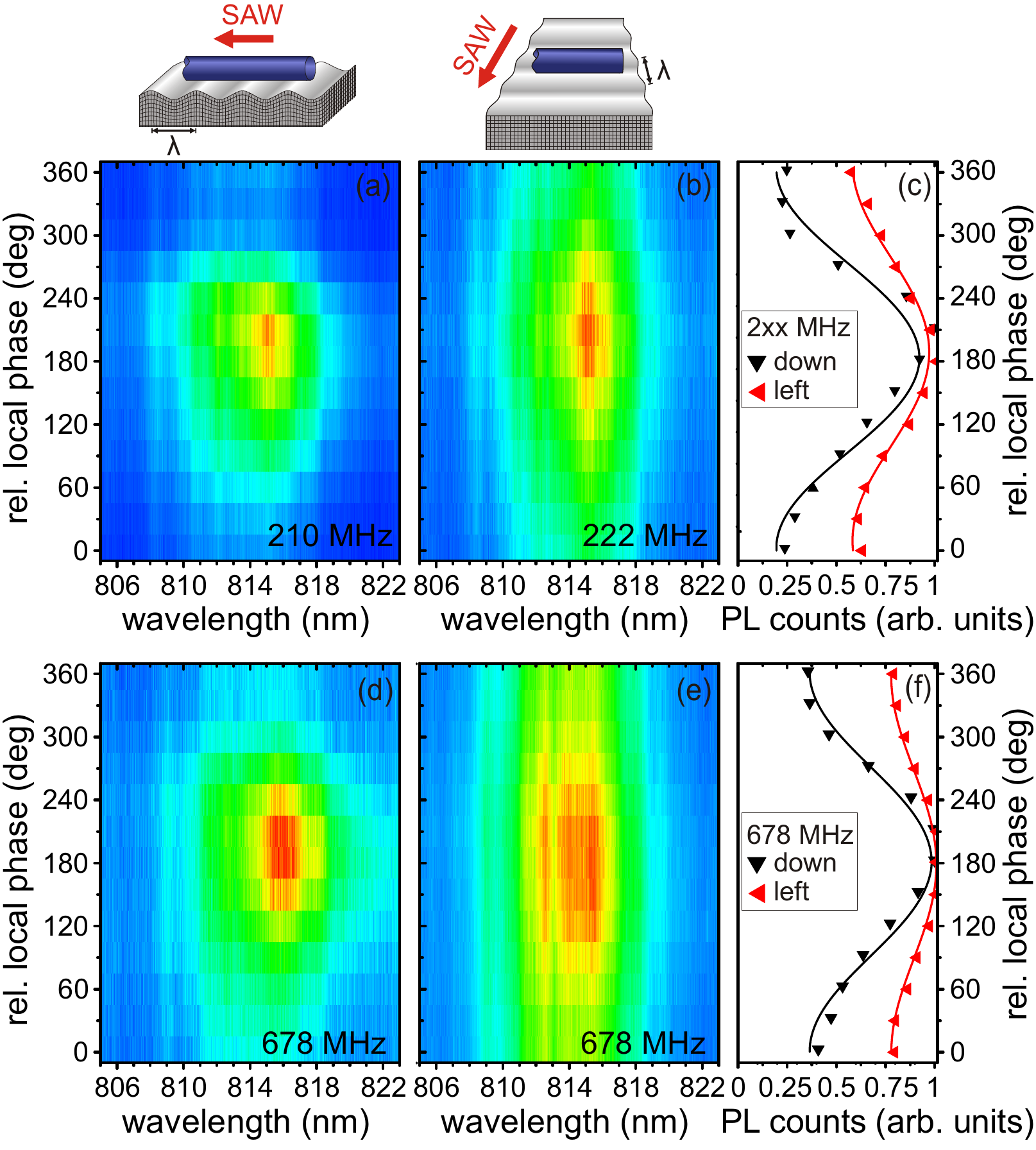}
        \caption{Phase-resolved PL spectra recorded from {\bf NW1} for different directions [schematics] and frequencies  [210/222 MHz (a) and (b); 678 MHz (d) and (e)] for $P_{\rm RF}$ = 27 dBm. The dynamic modulation by the SAW-field is resolved in the real spectra and the integrated intensities (c) and (f). Finite modulation for $\hat{k}_{\rm SAW}$$\perp$$\hat{a}_{\rm NW}$~arises from a slight misalignment from the ideal $90^{\rm o}$ configuration.}
        \label{fig:6}
    \end{center}
\end{figure}

In the time-integrated experiments presented in Figures \plainref{fig:1}--\plainref{fig:3} we averaged over all local phases $(\phi_{\rm SAW})$ of the SAW. By actively locking the $f_{laser}$ to the RF signal used to excited the SAW, we are able to perform phase resolved spectroscopy \cite{Voelk:11a} and resolve the dynamics of the emission modulation. In \ref{fig:6} we present normalized PL spectra as a function of $(\phi_{\rm SAW})$ plotted in false color representation. All spectra were recorded from {\bf NW1} and constant $P_{\rm RF}= +27{~\rm dBm}$ applied to the low frequency pair {\bf IDT1}-{\bf IDT2} [c.f. \ref{fig:6} (a) and (b)] and the high frequency pair {\bf IDT3}-{\bf IDT4} [c.f. \ref{fig:6} (d) and (e)], respectively. The relative orientation is $\hat{k}_{\rm SAW}$$\parallel$$\hat{a}_{\rm NW}$~and $\hat{k}_{\rm SAW}$$\perp$$\hat{a}_{\rm NW}$~in the left and center panels as indicated by the schematics. For both frequencies we observe a clear and pronounced modulation of the emission intensity as we tune $\phi_{\rm SAW}$ which \emph{directly} reflects the local and dynamic modulation of the NW emission. As expected from the results obtained by phase-averaged spectroscopy, the contrast of the observed modulation is significantly reduced for $\hat{k}_{\rm SAW}$$\perp$$\hat{a}_{\rm NW}$~[c.f. \ref{fig:6} (b) and (e)] compared to $\hat{k}_{\rm SAW}$$\parallel$$\hat{a}_{\rm NW}$~[c.f. \ref{fig:6} (a) and (d)]. We directly compare the amplitudes of these modulations by plotting the integrated PL intensity as a function of $\phi_{\rm SAW}$ in \ref{fig:6} (c) and (f) for the low and high frequency SAWs. The observed directionality is again reflected in the reduced amplitudes of the oscillation for $\hat{k}_{\rm SAW}$$\perp$$\hat{a}_{\rm NW}$~(left triangles) compared to $\hat{k}_{\rm SAW}$$\parallel$$\hat{a}_{\rm NW}$~(down triangles). Similar observations were also made for {\bf NW2} further confirming the directionality of the SAW-driven exciton dissociation in one-dimensional NWs.\\

\begin{table}[htb]
    \centering
     \caption{Relative dynamic suppression of emission $\eta$ extracted from data in \ref{fig:5} and \ref{fig:6}.}
     \label{tab:1}
     \begin{tabular} {lcccc}
      \textbf{NW1}, Phase-locking&\vline& 210 MHz/222 MHz &\vline& 678 MHz \\ \hline \hline

     $\hat{k}_{\rm SAW}$$~\parallel~$ $\hat{a}_{\rm NW}$ &\vline& $0.66\pm0.15$&\vline& $0.46\pm0.07$\\
     $\hat{k}_{\rm SAW}$$~\perp~$ $\hat{a}_{\rm NW}$ &\vline&$0.25\pm0.05$ &\vline& $0.13\pm0.05$\\ \hline\hline
      \textbf{NW2}, TCSPC &\vline&222 MHz &\vline& \\ \hline \hline
     $\hat{k}_{\rm SAW}$$~\parallel~$ $\hat{a}_{\rm NW}$ &\vline& $0.64\pm0.13$&\vline& $$\\ \hline \hline
        \end{tabular}
\end{table}
The experimentally measured modulations are well reproduced by least square fits of sine functions shown as lines in \ref{fig:6} (c) and (f) which we use to quantify the efficiencies of this process for the two configuration. As a figure of merit we define the relative dynamic suppression of the modulation
\begin{equation}
    \eta=\frac{I_{max}-I_{min}}{I_{max}+I_{min}}.
    \label{eq:1}
\end{equation}
The obtained values of $\eta$ for $\hat{k}_{\rm SAW}$$\parallel$$\hat{a}_{\rm NW}$~$(\eta_\parallel)$ and $\hat{k}_{\rm SAW}$$\perp$$\hat{a}_{\rm NW}$~$(\eta_\perp)$ are summarized for all four IDTs in \ref{tab:1}. From the ratio $\eta_\parallel/\eta_\perp$ we can quantify the efficiencies of the dynamic PL modulation for a constant RF power to 3.6 and 2.6 for the low and high frequency SAWs, respectively. Remarkably, $\eta_\parallel$ is almost identical for both frequencies which can be readily explained by the fast PL decay for uncapped, zinc blende phase GaAs NWs of $\tau_{PL}<30\mathrm{~ps}\sim 0.02\cdot f^{-1}_3 $ governed by surface recombination\cite{Demichel:10a} which is almost two orders of magnitude faster than the period of the SAW. Due to this fast decay the SAW represents a quasi-static perturbation and the observed modulation reveals the dependency of the SAW-driven exciton dissociation as a function of the local phase of the SAW. Moreover, a comparison of the length scale on which electrons and holes are separated ($\lambda_{3,4}/2=2.4 \mathrm{~ \mu m}$) and the diameter of the excitation laser focus ($\sim$3 $\rm \mu m$) we would be expect a more pronounced reduction of $\eta$ when changing the SAW frequency by more than a factor of 3 from $f_1= 210\mathrm {~MHz}$ to $f_3=678\mathrm {~MHz}$ \cite{Voelk:11a}. The weak but finite modulation and $\eta_\perp$ observed for SAWs excited by {\bf IDT2}  and {\bf IDT4} could arise from the non-perfect alignment of the NW as seen in the SEM image in \ref{fig:1} (a). Furthermore we want to note that $\eta$ measured on {\bf NW1} for $\hat{k}_{\rm SAW}$$\parallel$$\hat{a}_{\rm NW}$~[c.f. \ref{tab:1}] are in excellent agreement with $\eta = 0.64\pm0.13$ extracted from the TCSPC data recorded from {\bf NW2} shown in \ref{fig:5} (a) at the same RF power. \\

In summary we have demonstrated fully \emph{dynamic} and \emph{directional}, SAW controlled modulation of the optical emission of single \ce{GaAs} NWs. The underlying separation of electrons and holes provides the foundation for elaborate schemes to transfer individual charges and spins in axial\cite{Bjork:02,*Weert:09a,*Kouwen:10a} and radial\cite{Morral:08b,*Uccelli:10a} nanostructures defined in this inherently scalable architecture. To achieve such deterministic charge conveyance losses due to surface recombination which is the dominant mechanism of unpassivated NWs, have to be overcome e.g. by using surface-passivated core-shell NW structures\cite{Demichel:10a}. Moreover, a wide range of sophisticated architectures such as SAW-tunable crystal phase superlattices\cite{Spirkoska:09}, interdot couplings \cite{Krenner:05b,*Stinaff:06,*Robledo:08} and advanced memory devices \cite{Krenner:08a,*Heiss:09} or acoustically triggered generation of single photons  \cite{Borgstrom:05a,*Tribu:08a,Wiele:98,*Couto:09,*Voelk:10a} and polarization entangled photon pairs \cite{Singh:09a} can be realized.\\

\begin{acknowledgement}
This work was financially supported by DFG via the cluster of excellence {\it Nanosystems Initiative Munich} (NIM) and SFB 631, by BMBF within the EPHQUAM consortium and by the European Union via SOLID, the Marie Curie FP7 Reintegration Grant (Christina Totte, project officer), and the TUM Institute of Advanced Study (IAS). G.K. would further like to thank A. Fontcuberta i Morral and E. Uccelli for helpful discussions.
\end{acknowledgement}

\begin{suppinfo}
Description of molecular beam epitaxy growth and parameters, SAW device fabrication, gating scheme for time- and SAW phase-resolved optical spectroscopy.
\end{suppinfo}

\bibliography{report}

\end{document}